\documentclass{emulateapj}
\usepackage{lscape}

\newcommand{\myemail}{adamo@astro.su.se}

\newcommand{\msun}{M_{\odot}}

\slugcomment{Accepted for publication in ApJ; draft version October 18$^{th}$, 2010}

\shorttitle{The red excess in SBS 0335-052E}
\shortauthors{Adamo, A. et al.}

\begin{document}


\title{On the origin of the red excess in very young super star clusters: \\
the case of SBS 0335-052E}


\author{A. Adamo\altaffilmark{1},
E. Zackrisson\altaffilmark{1},
G. \"Ostlin\altaffilmark{1},
and M. Hayes\altaffilmark{2}}
\email{\myemail}

\altaffiltext{1}{Department of Astronomy, Stockholm University, Oscar Klein Center, AlbaNova, Stockholm SE-106 91, Sweden}
\altaffiltext{2}{Observatoire Astronomique de l'Universit\'{e} de Gen\`{e}ve, 51, ch des Maillettes, CH-1290 Sauverny, Switzerland}



\begin{abstract}
 The spectral energy distribution analysis of very young unresolved star clusters challenges our understanding of the cluster formation process. Studies of resolved massive clusters in the Milky Way and in the nearby Magellanic Clouds show us that the contribution from  photoionized gas is very important during the first Myr of cluster evolution. We present our models which include both a self-consistent treatment of the photoionized gas and the stellar continuum and quantify the impact of such nebular component on the total flux of young unresolved star clusters. A comparison with other available models is considered. The very young star clusters in the SBS 0335-052E dwarf starburst galaxy are used as a test for our models. Due to the low metallicity of the galactic medium our models predict a longer lasted nebular phase which contributes between 10-40 \% of the total near infrared (NIR) fluxes at around 10 Myr. We propose thus a possible solution for the observed flux excess in the 6 bright super star clusters of SBS 0335-052E. Reines et al.  showed that the observed cluster fluxes, in the red-optical and NIR range, sit  irreconcilably above the provided stellar continuum models. We find that in the age range estimated from the H$\alpha$ emission  we can explain the red excess in all the 6 super star clusters as due to nebular emission, which at cluster ages around 10 Myr still affects the NIR wavebands substantially. 
\end{abstract}



\section{Introduction}
\begin{figure*}
\plotone{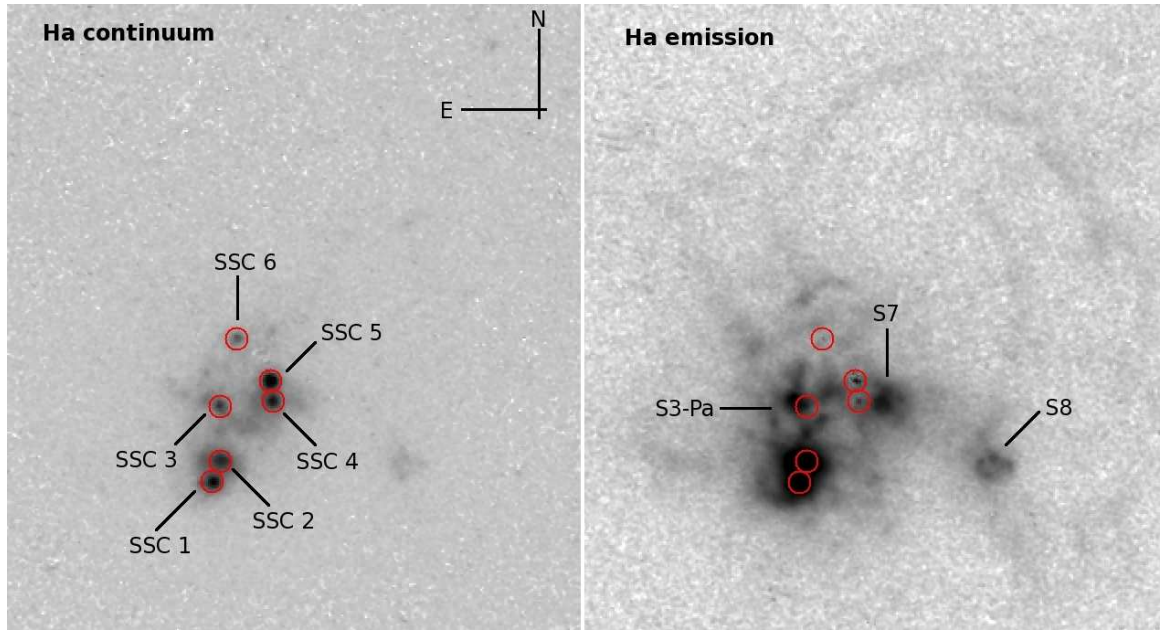}
\caption{H$\alpha$ continuum and emission frames of SBS 0335-052 \citep{2009AJ....138..923O}. The position of the compact SSCs is showed by the red circles. The radius of the circles is 0.15". The new star forming regions with bright H$\alpha$ emission are indicated in the right frame using the nomenclature by T09.}
\label{halpha}
\end{figure*}

SBS 0335-052E, at a distance of 54 Mpc (according to NED), is one of the least chemically evolved galaxies in the local
universe (Z=1/40 Z$_{\odot}$, \citealp{1990Natur.343..238I}; \citealp{1997ApJ...476..698I}; \citealp{2000A&ARv..10....1K};  \citealp{2006A&A...454..119P}). In the 1990s, the first {\it Hubble Space Telescope} ({\em HST}) images of this irregular dwarf galaxy \citep{1997ApJ...477..661T} revealed that the main starburst region was dominated by 6 young super star clusters (SSCs). In Figure~\ref{halpha} we show the position of the clusters as identified by \citet{1997ApJ...477..661T} using the nomenclature of \citet[][hereafter R08]{2008sbc0335R}. The clusters are located around a supernova cavity which has probably caused their formation, with star formation propagating from north to south \citep{1997ApJ...476..698I}. 

\citet[][hereafter T09]{2009ApJ...691.1068T} observed bright Pa$\alpha$ emission in the galaxy and identified three new and actively star forming regions, S3-Pa$\alpha$, S7 and S8 (indicated in the continuum subtracted H$\alpha$ image in Figure~\ref{halpha}). These new areas were presented as places of less clustered star formation, where young stars are ionizing the gas in situ. Similar star forming regions were also observed around SSC1 and SSC2.

\citet{2000A&A...363..493V}, found that a hot dust component is needed to explain the red $H-K$ color observed globally in the galaxy. However, the very low visual extinction across the SSCs, which show prominent  Br$\gamma$ emission, implies that the clusters are quite dust free. Interestingly, \citet{2001A&A...377...66H}, using imaging and spectroscopy data at $2-4$ $\mu$m, found evidence of a deeply embedded cluster with a visual extinction $A_V\geq15$ mag in the line of sight of the two young SSC1 and 2. They excluded that the IR flux was produced by the two optical bright SSCs because of the high discrepancy between the extinction derived from the IR and from the optical data ($A_V \sim 0.55$ from the optical line ratios). 

A different scenario was presented by R08 and \citet{2009AJ....137.3788J} to reconcile the apparently discrepant extinction
estimates in the knot containing SSC1 and 2. R08 inferred $\sim 40$\% of Lyman continuum photon leakage from the two clusters which is ionizing the extended H$\alpha$-bright region around the knot.  This could be possible if the  medium surrounding the two clusters is clumpy. In this case the optical light would come through the "holes" suffering only low extinction, while in the NIR would transmit also the signal produced by the dense dust clumps, heavily extinguished. 

Using H$\alpha$ equivalent width EW(H$\alpha$) measurements, R08 deriveded cluster ages between 3 and 15 Myr. The 6 SSCs also displayed a flux excess at wavelengths larger than 8000 \AA, \ impossible to reconcile with the stellar continuum models used to fit the cluster spectral energy distribution (SED). R08 explored several possibilities to explain the origin of the red excess. The excess at $\sim 0.8 \mu$m was explained as the dust photoluminescence phenomenon commonly referred to as extended red emission (ERE, see \citealp{2004ASPC..309..115W}). The ERE has been observed in star forming regions like 30 Doradus \citep{1998A&A...333..264D} and in H{\sc ii} regions of local galaxies like NGC 4826 \citep{2002ApJ...569..184P}, and is characterized by an extended emission feature between 7000 and 9000 \AA. R08 attributed the flux excess at wavebands between 1.6 and 2.1 $\mu$m to two different mechanisms. For the youngest clusters SSC1 and SSC2,  hot dust ($\sim800$ K) emission was advocated to explain the rise of the IR continuum. The remaining older clusters showed infrared colors consistent with those of red super giants (RSGs). In fact, if the models do not contain a reliable treatment of the RSGs (\citealp{2008A&A...486..165L} and references therein) this could produce an apparent NIR excess in the observed SEDs at cluster ages around 10 Myr. 

The clusters are very young ($3-15$ Myr). This age range is quite complex, due to the rapid dynamical evolution that the clusters experience (see \citealp{2010arXiv1002.1961P} for a review).  At such young stages, the contribution from photoionized gas in the broadband integrated fluxes can be substantial ( \citealp{2002A&A...390..891B}; \citealp{Anders & Fritze-Alvensleben}; \citealp{Zackrisson et al. b}; \citealp{R2009}). In the starburst galaxy Haro 11, we found a peak  of cluster formation only 3.5 Myr old  \citep{Adamo10}. The analysis of such young clusters requires models that include both the stellar continuum and photoionized gas emission contributions (see \citealp{Zackrisson et al. a}, and \citealp{Adamo10}).  However, we also found that a subsample of clusters in Haro 11 dysplays a flux excess above models including photoionized gas emission at wavelengths longward of 8000 \AA. The observed discrepancies could be explained if other mechanisms, not included in the models and related to the young ages of the clusters, are contributing at redder wavebands. Hot dust, a considerable fraction of pre-main sequence stars (PMSs) and young stellar objects (YSOs), are possible mechanisms causing the red excess in Haro11. Previous cases of NIR excess observed in still embedded star clusters of dwarf starburst galaxies was explained by several authors as caused by hot dust (\citealp{2005A&A...433..447C}; \citealp{2005ApJ...631..252C}). Observations of very young star clusters in the Small Magellanic Clouds have revealed complex environments composed by bright main sequence stars, but also pre-main sequence objects (PMS) and young stellar objects (YSOs) only observable in the NIR  (\citealp{2007ApJ...665L.109C}; \citealp{2010A&A...515A..56G}). A significant fraction of YSOs was suggested as the origin of the red excess in the starburst galaxy NGC 253 \citep{2009MNRAS.392L..16F}. 

The NIR excess in young star clusters has been observed in several nearby starburst galaxies. Here, we explore to what extent the red excess reported in SBS 0335-052E by R08 can be explained by nebular emission, or other mechanisms are needed.

In Sections \ref{mod} and \ref {sec:mod} we present the photometric data and our synthetic evolutionary models which we use to perform new fits to the cluster SEDs as described in Section \ref{sec:sed}. In this section we also test the validity of our fits and attempt an estimation of the sizes of the emitting H{\sc ii} regions. In Section \ref{disc} we compare our results with previous studies of the galaxy. Our conclusions are summarized in Section \ref{conc}.

\section[]{Data sample}
\label{mod}
The data reduction and photometry were carried out and described in detail in Section 3 of R08. In the present work we will use their photometry (published in their Table 1) to test our models and investigate the origin of the red excess in SBS 0335-052E.

In Figure~\ref{halpha} we show two of the science frames of SBS 0335-052E, previously published by \citet{2009AJ....138..923O}. We illustrate the position of the clusters as in R08 and the star forming regions identified by T09. The photometric annuli around the clusters have a radius of 0.15" ($\sim$40 pc at the adopted distance). This corresponds to the size set by R08 to perform aperture photometry. The six SSCs are bright and visible in all the UV, Optical and IR {\it HST} frames. The continuum frame in Figure~\ref{halpha} shows that the positions of the compact clusters are clearly defined and that the used aperture size is suitable to avoid contamination from the neighboring diffuse regions. On the other hand, in the right frame, we see that H$\alpha$ emission is quite strong at the location of the SSC1 and SSC2  which makes it difficult to distinguish between the compact and the less clustered surrounding  regions.  The remaining 4 star clusters do not show H$\alpha$ fluxes as strong as the ones produced in the nearby very young active star forming regions. 

\subsection{New estimation of the EW(H$\alpha$) in the clusters.}
\label{ew_ha}

 The use of local sky annuli, like the ones used by R08, could oversubtract flux from the sources, producing an underestimation of the final EW(H$\alpha$). Due to the complexity of the environment around the clusters, we decided to re-estimate the EW(H$\alpha$) across the clusters using the same aperture radius as R08, but a constant value for the background as explained below. 

The reduction of science frames shown in Figure~\ref{halpha} and software are described in more detail in \citet {2009AJ....138..911H}, here, we briefly summarize the procedure. The continuum subtraction of the H$\alpha$\ frame ({\em FR656N}) is non-standard, and utilizes custom software. This software performs spatially resolved, pixel-by-pixel SED fitting using five {\em HST} broadband filters between the FUV and optical domain, all of which were expressly chosen to be free from contamination by nebular lines. From the best-fitting synthetic stellar population, it computes the expected flux due to continuum processes in the {\em FR656N} narrowband filter. This is then subtracted from the observation itself, and maps of the pure 6563~\AA\ continuum, and continuum-subtracted H$\alpha$\  are output. 

Photometry was carried out in both frames, at the same position, using the same aperture radius (0.15") as in R08. Because the H$\alpha$ emission has a complex and extended morphology, we preferred to subtract a constant mean value of the sky background instead of a local sky annulus located around the position of the cluster, in both continuum and emission frame. The mean sky values were estimated in both frames, in the same position, and in a region that appeared  free of H$\alpha$ emission and continuum flux. No aperture corrections to the determined fluxes were applied since there are no point-like sources in the reduced frames and simulated PSFs  would introduce additional uncertainties. For this reason, it is possible that the new estimated EWs could be a lower limit to the real values. However, the EW is a ratio, so if line and continuum have the same distribution (an assumption that is valid for high EWs) the aperture correction would be cancelled. 

The EWs were estimated by the ratio between the H$\alpha$ emission flux multiplied by the filter width and the flux in the continuum. In Table~\ref{age-halpha}, measured H$\alpha$ flux and EWs are listed for all the clusters. In general, the observed EW(H$\alpha$) found by R08 are smaller than the ones estimated here. The main differences between the two analyses reside in the reduction of the frames, and in the performed sky background subtraction. The continuum frame, in our analysis, is better constrained than the one produced by a simple interpolation between the two nearest frames to H$\alpha$. 

\begin{figure*}
\centering
\includegraphics[scale=0.8]{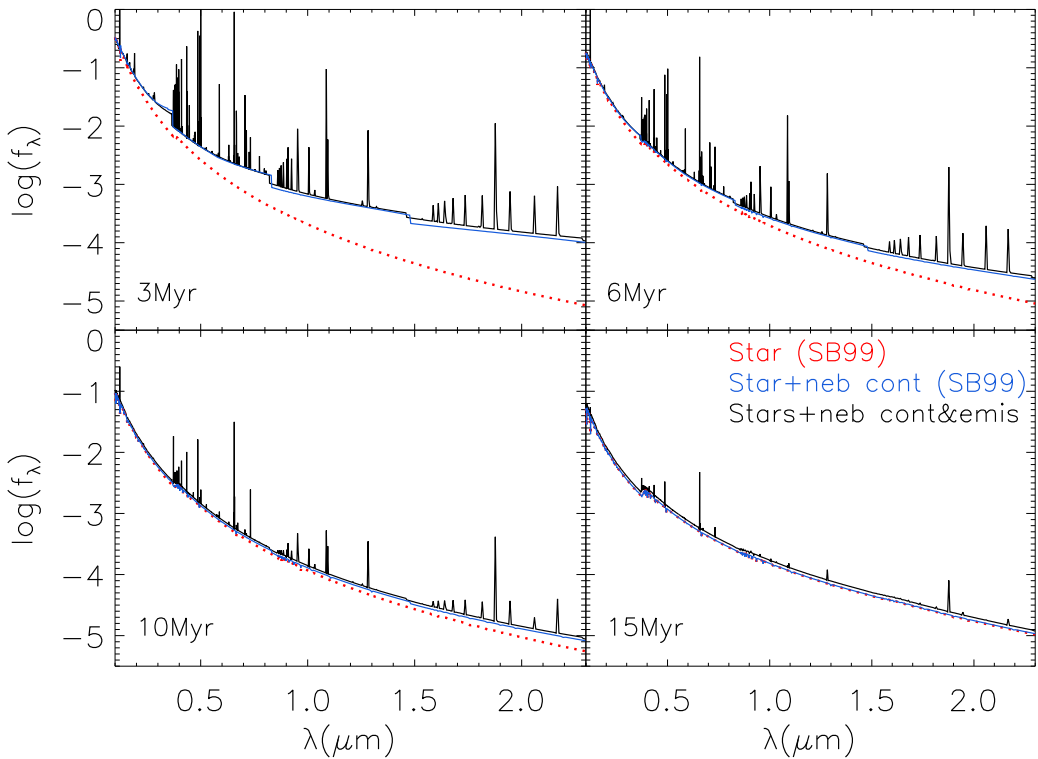}
\caption{Evolution of the spectral model for a $10^6 \msun$ cluster. In black we show the model including a full treatment on the nebular emission and continuum, in red the stellar continuum model, and in blue the stellar and nebular continuum only (the last two models are outputs from Starburst99). The fluxes are in arbitrary units. The wave range is between 0.1 and 2.3 $\mu$m. }
\label{model}
\end{figure*}

\section{Description of the spectral evolutionary models}
\label{sec:mod}

\begin{figure}
\includegraphics[scale=0.48]{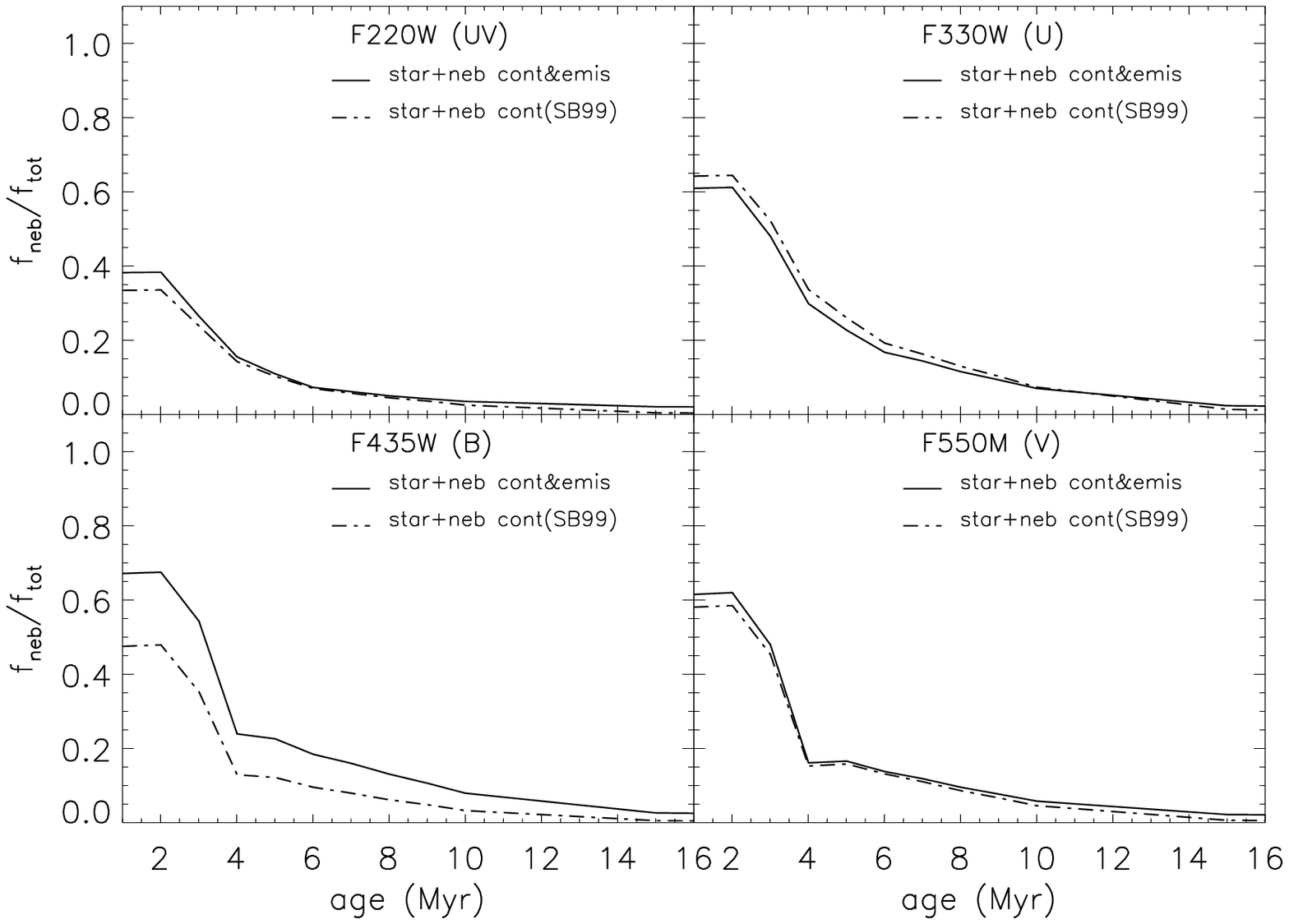}
\includegraphics[scale=0.48]{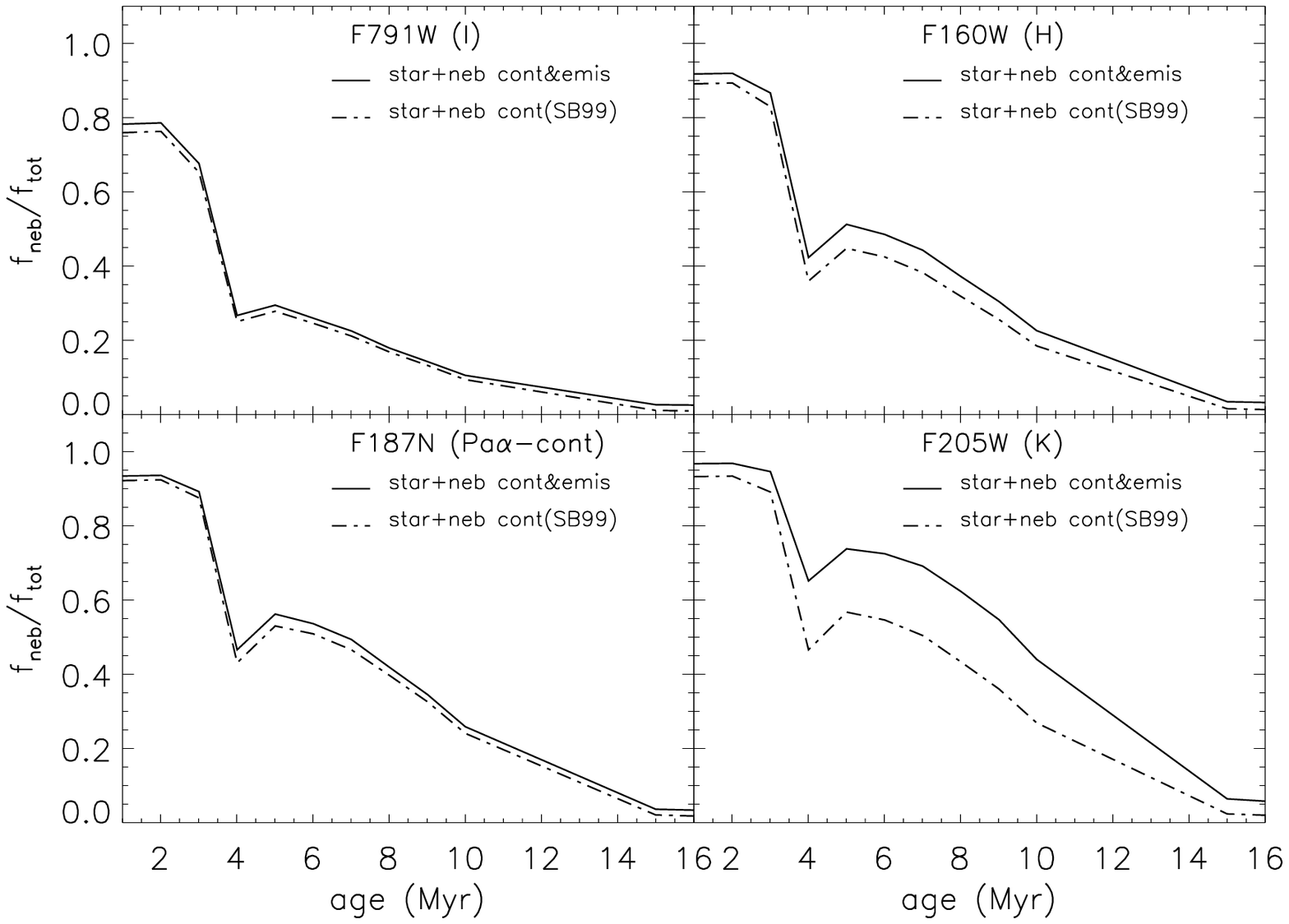}
\caption{Relative contribution  of the nebular emission as function of the cluster age to the total flux transmitted in the {\em HST} filters used in this analysis. The SB99 models including stellar and nebular continuum are showed with a dashed dotted line. Our models are drawn by solid lines.  The filters are indicated in the plots. A description of the models is given in section \ref{sec:mod}.}
\label{model2}
\end{figure}

To produce self-consistent models which include the contributions from both stars and photoionized gas, we use the stellar population SEDs predicted by Starburst99 \citep{Leitherer et al.} as input to the Cloudy photionization code \citep{Ferland et al.}. The procedure closely follows that described by \citet{Zackrisson et al. a}, in which Cloudy is called for every stellar population time step to ensure that the temporal evolution of the nebulae is treated self consistently.  This technique results in combined SEDs that are more realistic than the ones generated when using the nebular components produced by Starburst99 itself. The latter contain only nebular continuum (produced by free-free and free-bound emission), whereas the models we use  include both nebular continuum and emission lines (produced by bound-bound mechanisms). In our models, the SB99 stellar population SEDs are produced using the Padova tracks \citep{1994A&AS..106..275B} and assume a metallicity of $Z=0.0004$, an instantaneous burst of star formation (resulting in a single-age population) and a stellar initial mass function with a double power-law ($\mathrm{d}N/\mathrm{d}M \propto M^{-\alpha}$, with $\alpha=1.3$ for $M=0.1$--$0.5\ M_\odot$ and  $\alpha=2.3$ for $M=0.5$--$100\ M_\odot$). The nebular SED assumes a spherical, constant-density and ionization-bounded nebula with a gaseous metallicity identical to that of the stars. The gas filling factor $f$ (which quantifies the porosity of the nebula), the hydrogen number density $n_H$, and the gas covering factor $c$ (which quantifies the amount of Lyman continuum leakage due to clumpiness in the medium or presence of dust) are all explored within reasonable boundaries. We generated a grid of models, with different combinations of values for $f=[0.01, 0.001]$, $n_H = [10^2, 10^4]$ cm$^{-3}$, and $c=[0.1, 0.2, 0.3, 0.4, 0.5, 1.0]$ ($c=1$ means that all the Lyman continuum photons ionize the gas; $c=0.1$ that only the 10 \% of them photoionizes the gas, i.e. that 90 \% of photons are leaking).

In the Figure~\ref{model}, we show how the integrated spectrum of a $10^6 \msun$ cluster changes as a function of the age during the first 15 Myr, using our models with $f=0.01$, $n_H=10^4$ cm$^{-3}$, and $c=1$.  For comparison  we plot the corresponding Starburst99 stellar continuum spectrum (red dotted line) and the model with nebular continuum included (blue line). In the very early stages,  around 3 Myr, the contribution from the photoionized gas is so strong that the whole spectral range from the near-UV to IR is affected. The stellar continuum is significantly below. The continua of the two models including the nebular contribution are in good agreement. However our models, implemented with Cloudy,  include also emission lines, very strong at this age. At 6 Myr, we observe that the stellar continuum and the spectrum of the model including both stellar and nebular components are similar in the near-UV and blue-optical range. The contribution from the line emission, however, is very important in the Balmer region. From the red-optical to the near-IR range the difference in fluxes between models with and without nebular treatment is still substantial. Around 10 Myr, fluxes at wavelengths longward 1.0 $\mu$m  still receive a non negligible fraction of flux from the gas. This is evident using both SB99 (stellar and nebular continuum) and our models. At shorter wavelengths, only the integrated fluxes across the emission lines are affected. The nebular component is almost gone at around 15 Myr, with very small differences between the 3 spectra. 

In Figure~\ref{model2} we compare the fraction of nebular emission contributing to the total flux as estimated by our models and by the SB99 ones. We quantify these fractions as a function of the cluster age for the {\em HST} filters used in the analysis of SBS 0335-052E (R08). As already showed in the previous figure, the only significant difference between the two models is the presence/absence of emission lines. On the top panels, we clearly see the drop in the contribution of the photoionized gas  already at $4-6$ Myr in the NUV ({\it ACS}/F220W), U ({\it ACS}/F330W), B ({\it ACS}/F435W), and V ({\it ACS}/F550M) filters. During this age range, the difference between the two models is of $\leq 5$ \% and disappears at older ages, if the filter does not transmit important lines (i.e. NUV, U, and V). On the other hand, the B filter shows in our models a significantly higher fraction of photoionized gas contribution, due to the presence of the Balmer lines. Around 10 Myr the B band still transmits 10 \% of the nebular flux which is mainly contained in the lines (see Figure~\ref{model} at $\sim 0.4-0.45$ $\mu$m).  In the remaining UV and optical bands, only 5 \% of the integrated flux is produced by the nebula. In the bottom panels we show the NIR filters. Again the discrepancy between our models and the SB99 ones is higher in filters which transmit strong lines (mainly in K). As already noticed before in the NIR range, at 10 Myr, the nebula still supplies 10 \% of the total flux in the optical I ({\it WFPC2}/F791W) band, 20 \% in the H ({\it NIC2}/F160W) band and in the Pa$\alpha-$continuum narrow filter ({\it NIC2}/F187N), and up to 40 \% in the K ({\it NIC2}/F205W) band. We also notice that at a given age the fraction of flux produced by the ionized gas increases as function of the wavelength, and shows a slower decline towards older ages.

\begin{figure}
\includegraphics[scale=0.48]{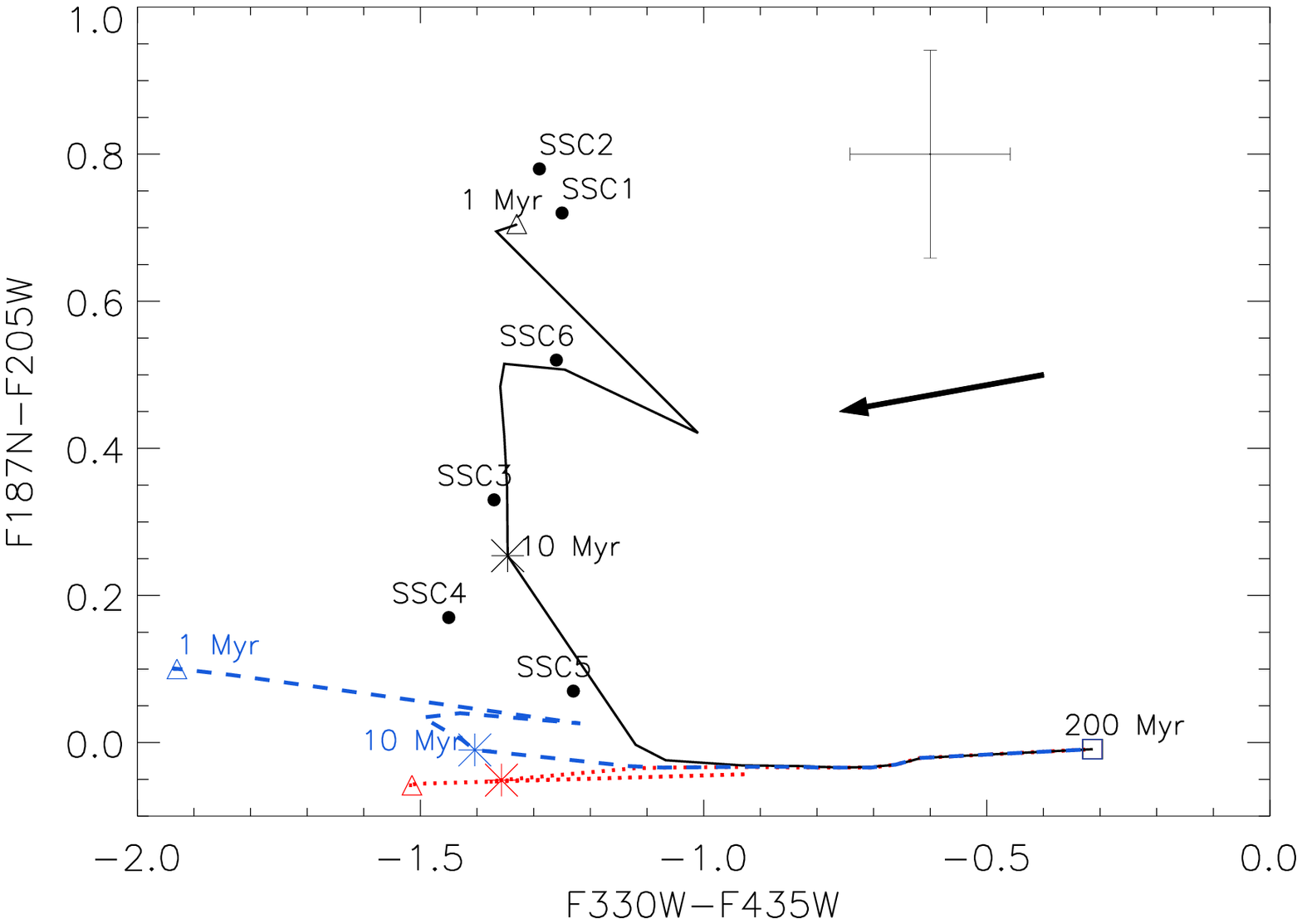}
\includegraphics[scale=0.48]{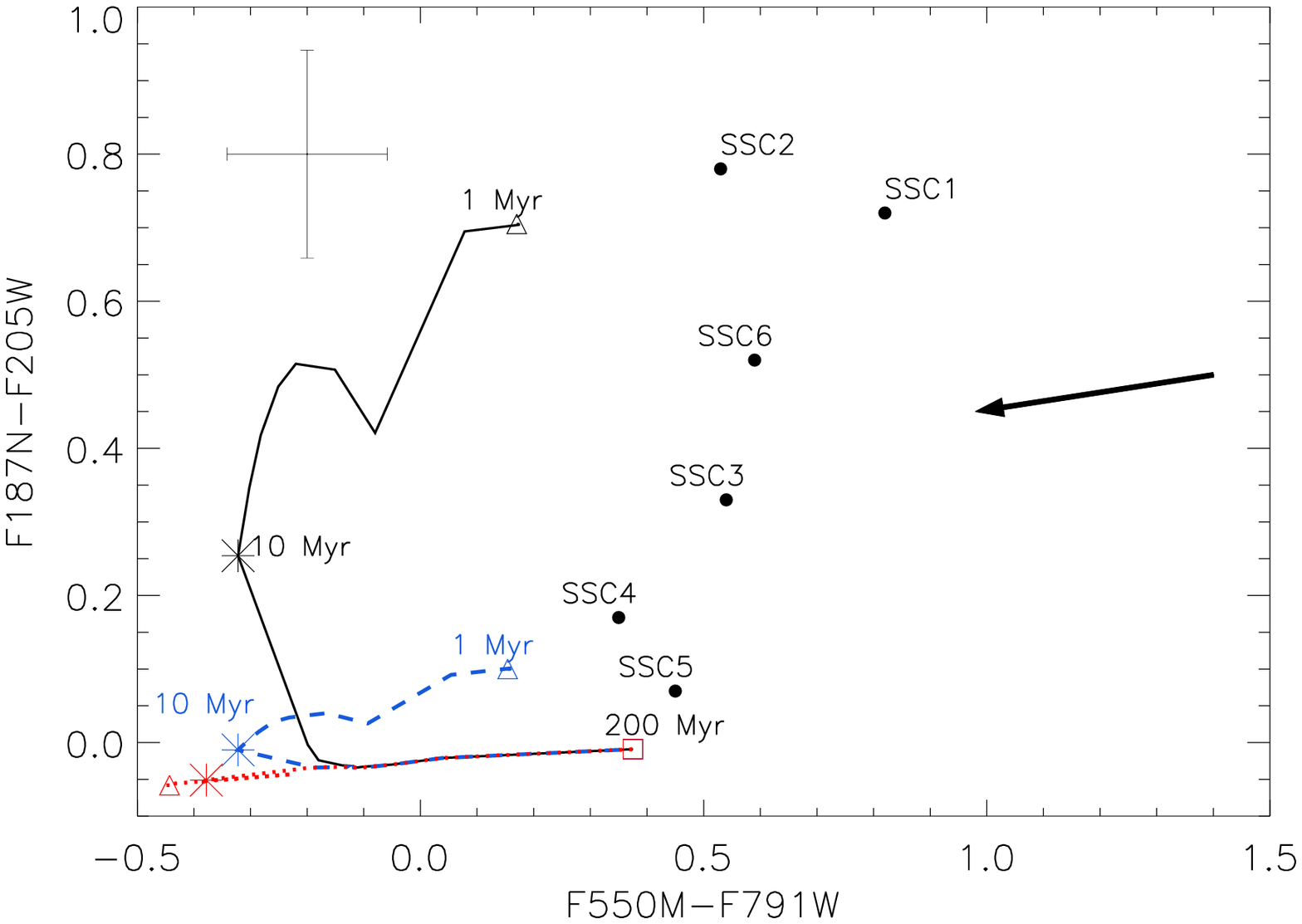}
\caption{The $F187N-F205W$ color versus the $F330W-F435W$ and the redder $F550M-F791W$ ones in two color diagrams. The evolutionary track provided by our model is represented by a solid line. We included Starburst99 evolutionary tracks with only stellar continuum (red thick dotted line) and stellar plus nebular continuum (blue thick dashed line) as well. Positions in the tracks corresponding to different ages are labelled. The colors of the six SSCs in SBS 0335-052E are overplotted. The arrow shows in which direction the correction for an visual extinction of A$_V=1.0$ mag moves the position of the clusters in the two diagrams. Photometric error bars are included.}
\label{CCD}
\end{figure}

As further test, in Figure \ref{CCD} we plot the evolutionary tracks in different combination of IR and optical colors. As already observed in \citet{Zackrisson et al. a}, during the first 10 Myr the predicted colors by models which include contribution from photoionized gas are quite different from the ones with only stellar continuum. We also plot, in the color diagram, the position of the six SSCs of SBS 0335-052E. We clearly see that the cluster IR colors are much redder than the predictions provided by the stellar continuum only or the stellar and nebular continuum tracks (SB99), but in fairly good agreement with our models. The position of the clusters in the bottom panel is shifted toward red F550M-F791W color ($V-I > 0$). This displacement is due to a small flux excess in I band, which is around $0.2-0.4$ mag above the best fitting models. This issue will be discussed in the following sections.

\section{Constraining the origin of the red excess}
\label{sec:sed}
\subsection{A new cluster SED fit from the NUV to the NIR}

The SED fit made by R08 was limited to the near-UV and blue-optical filters ({\it ACS}/F220W,  {\it ACS}/F330W,  {\it ACS}/F435W,  {\it ACS}/F550M). The Starburst99 spectral evolutionary models, including only the stellar continuum component were used together with the LMC extinction curve (\citealp{1999ApJ...515..128M}, and references therein) to estimate ages, masses and extinctions of the clusters. 

In Section \ref{sec:mod}, we argued that the photoionized gas is a crucial component for the models used to estimate the physical properties of the young clusters. In this section we explore the possibility to explain the red excess observed by R08 using these new models. The young clusters in SBS 0335-052E are indeed in the the age range where the contribution from the photoionized gas surrounding the clusters is expected to be important. The amount of flux produced by the gas and contributing to each of the cluster integrated fluxes is a function of the total amount of ionizing UV-flux produced by the massive stars in the cluster. The EW(H$\alpha$) is a direct measurement of the strength of the ionizing flux and very sensitive to the rapid evolution of the massive stars during the first 15 Myr of the life of the cluster. R08 compared the measurement of the EW(H$\alpha$) with predictions made by SB99 to derive the ages of the clusters.  We performed a similar estimation using the new measurements of EW(H$\alpha$) of each SSC  and the predictions provided by Cloudy in our models. The ages constrained by R08 and ours are reported in Table \ref{age-halpha} (fourth and fifth column, respectively). The ages we derived are in good agreement with what already found by R08 and by \citet{2009AJ....137.3788J} for the two bright radio sources SSC1 and 2. 

\begin{figure*}
\includegraphics[scale=0.48]{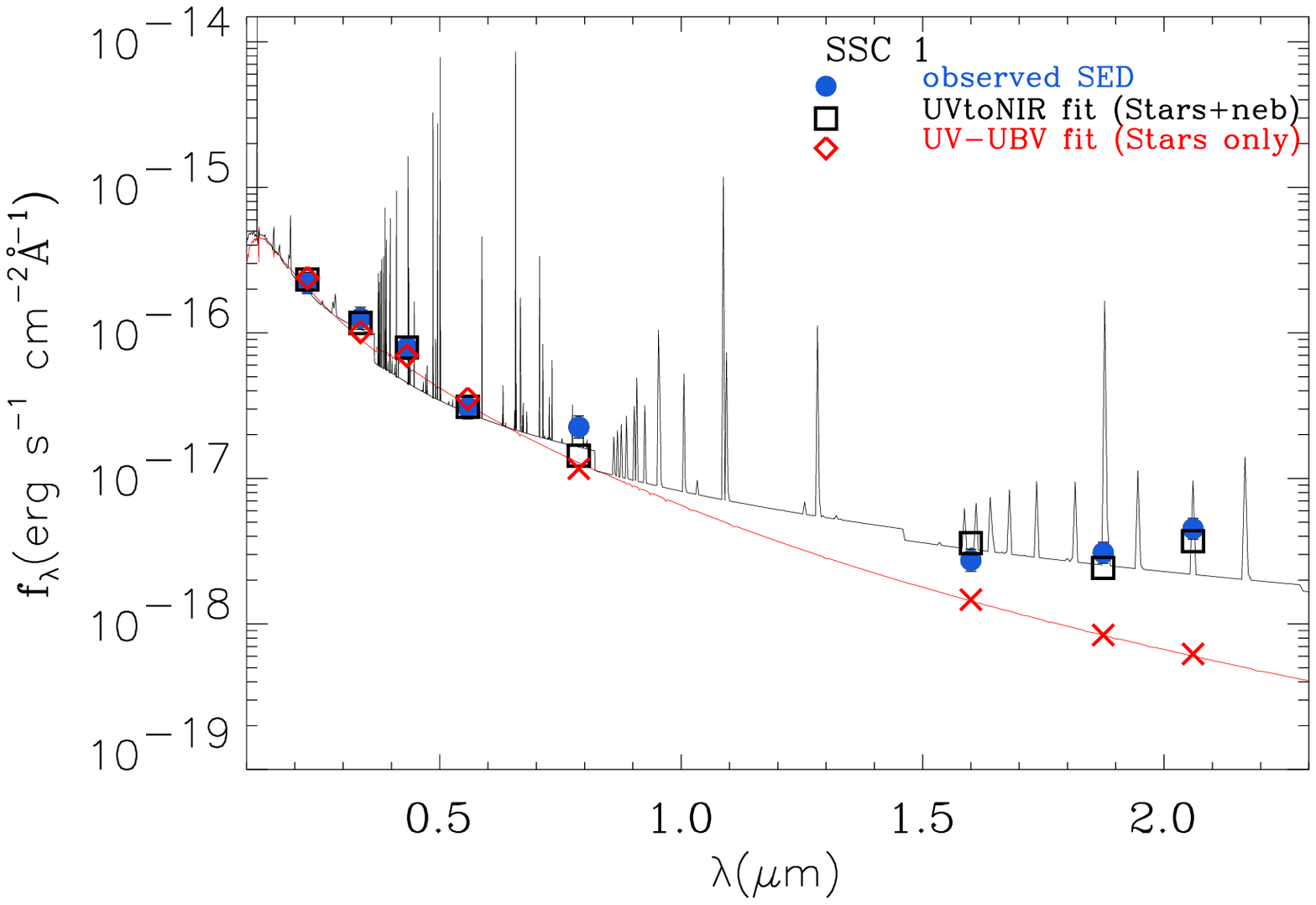}
\includegraphics[scale=0.48]{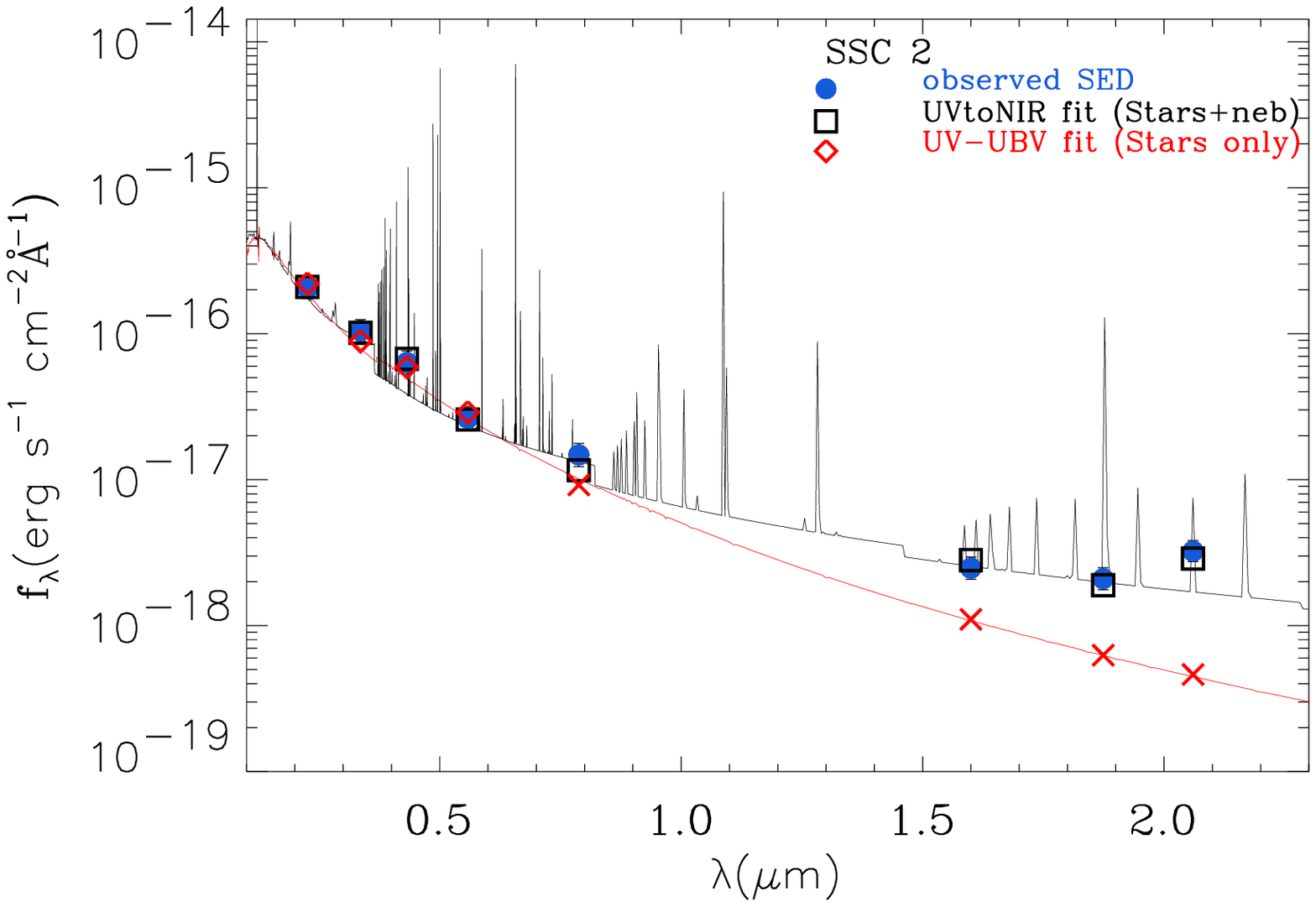}\\
\includegraphics[scale=0.48]{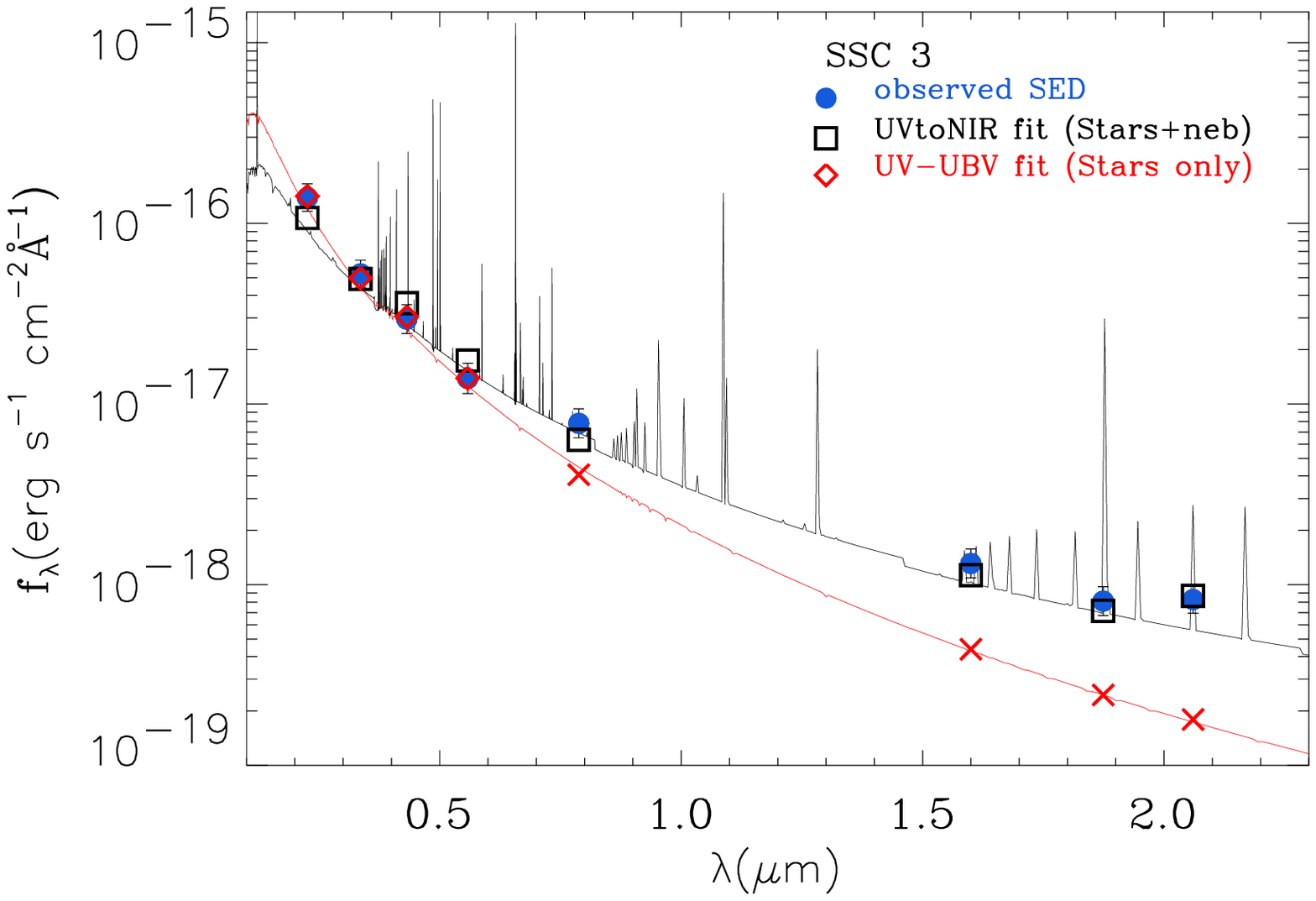}
\includegraphics[scale=0.48]{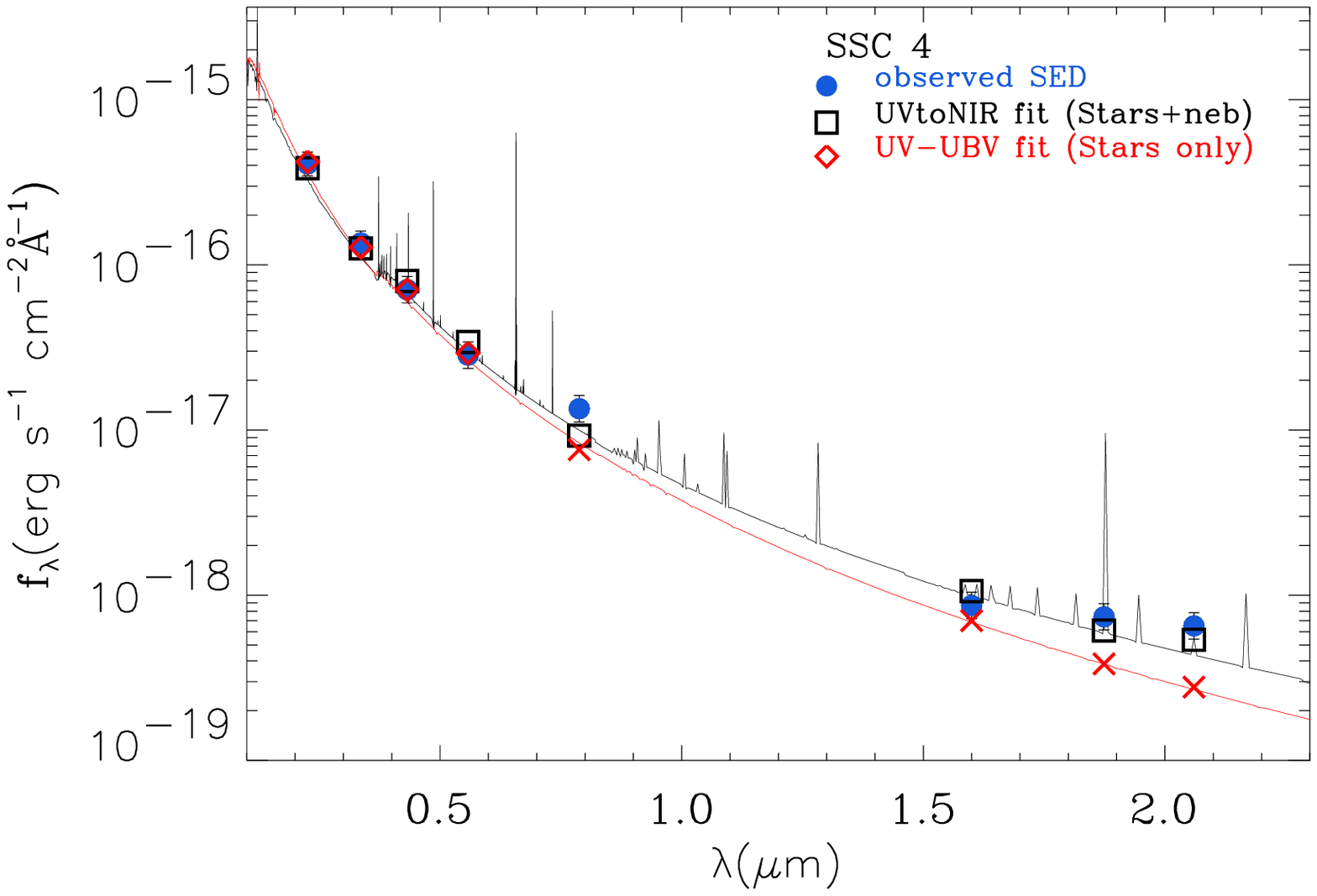}\\
\includegraphics[scale=0.48]{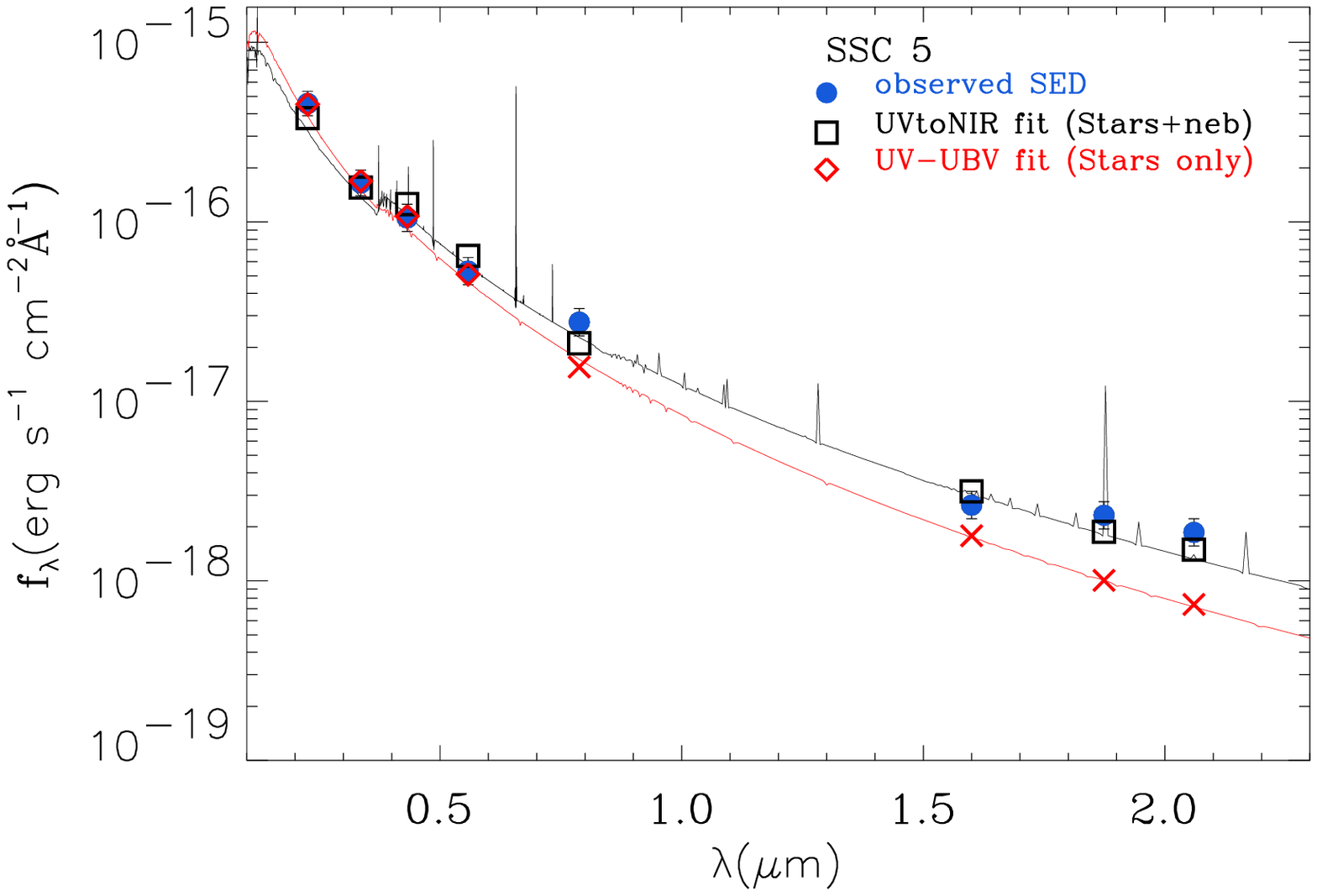}
\includegraphics[scale=0.48]{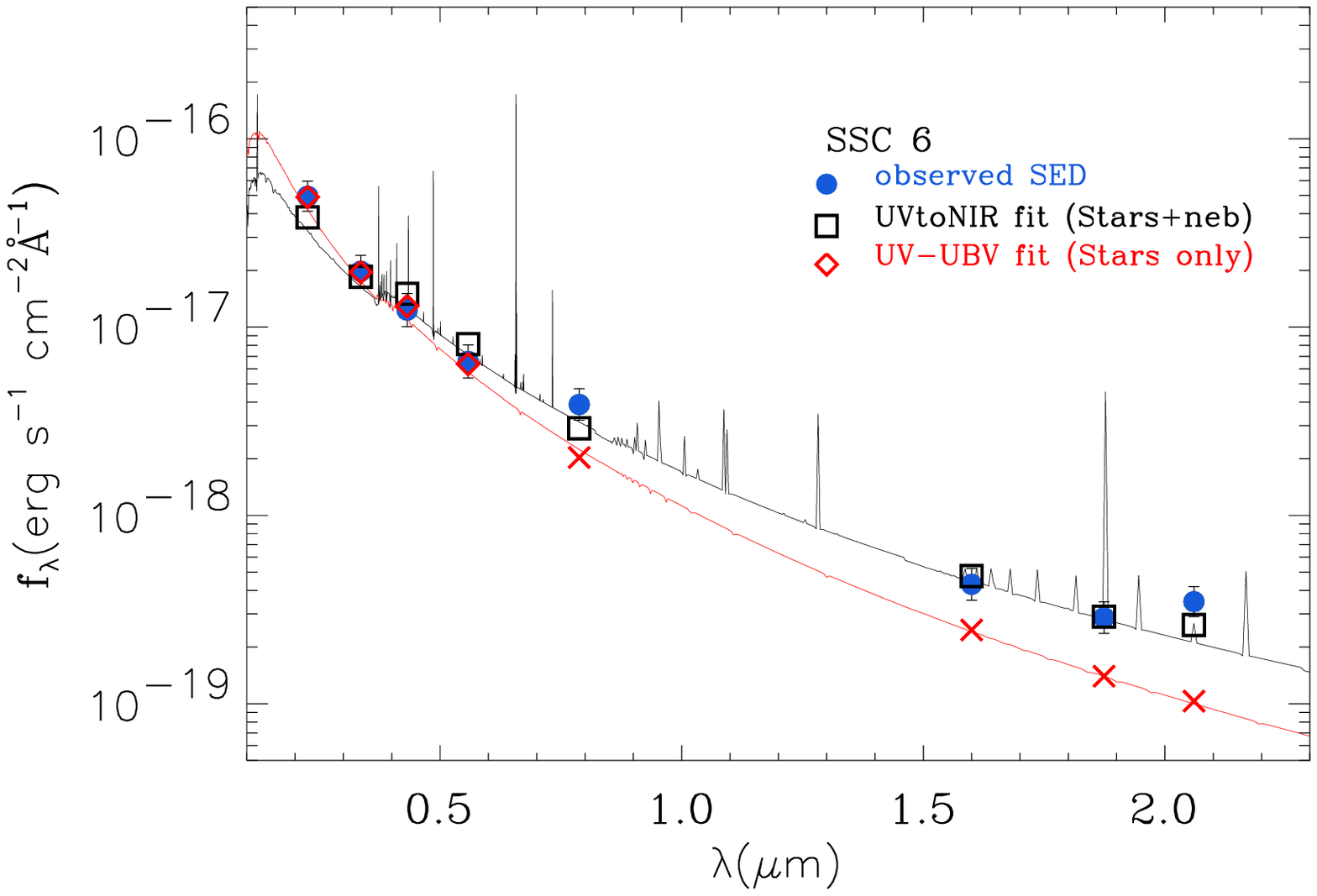}\\
\caption{SED analysis of the 6 clusters. The filled blues points are the observed integrated fluxes and associated error bars (R08). The black squares show the best fitted integrated fluxes produced by the our models. We plotted also the corresponding spectrum in black. In red we show the best fit produced by stellar continuum models. The red diamonds show the data points which have been included in the fit. The red crosses show the integrated fluxes excluded from the fit. In all the cases modeled spectra and integrated fluxes have been corrected for the estimated extinction. See the text for more details.}
\label{spec}
\end{figure*}

Keeping fixed the age, we produced a fit of the observed integrated fluxes of each cluster from UV to IR bands with the fluxes provided by our models at the corresponding ages. Our fit algorithm is described in detail in \citet{Adamo10}. Internal extinction was treated as a free parameter, and we used the Calzetti attenuation law \citep{2000ApJ...533..682C}, allowing extinction varying from $E(B-V)=0.0$ to 3.0 with a step of 0.01. In the previous fit, R08 used the LMC extinction law. We showed however, in \citet{Adamo10}, that there are not substantial deviations in the estimated ages, masses, and number of clusters affected by a red excess, when a Calzetti or a LMC law are used. The mass of the clusters was derived from the normalization between best-fitted model and observed data. The models that produce smaller residuals are the ones with: $f=0.01$, $c=1$, and $n_H=10^4$ cm$^{-3}$. The fits are shown in Figure~\ref{spec}, and the masses and A$_V$ are given in the Table \ref{age-halpha}. We noticed, however, that the models with same $f$ and $c$ values but lower hydrogen densities (n$_H = 10^2$ cm$^{-3}$) produce fits with residuals only slightly bigger than the best models.  The constrained extinction range is in good agreement with the values found using optical and NIR line ratio (e.g. \citealp{2000A&A...363..493V}). The values are also in agreement with the recovered extinction ranges in other starburst environments like the Antennae system \citep{2005A&A...443...41M} and M51\citep{2005A&A...431..905B}. The estimated cluster masses for the two youngest clusters, SSC1 and 2, are significantly lower than the ones determined by R08. They are instead, in agreement with the estimates by \citet{2009AJ....137.3788J} derived from the thermal radio emission of the two young SSCs, under the assumption of no leakage of ionizing photons from the nebula surrounding the young cluster.  The remaining clusters have estimated masses in agreement with the values previously obtained by R08.  During the first few Myr, where the photoionized gas emission is dominant, we clearly see that the cluster mass can be overestimated by a factor of $2-3$, if the used models do not include nebular contribution.
\begin{deluxetable*}{rccccccr}
\tablecolumns{8}
\tablewidth{0pt}
\tablecaption{Properties of the 6 SSCs. SED at fixed age for mass determination. Age is determined through  EW(H$\alpha$). \label{age-halpha}}
\tablehead{
\colhead{Ids} &
\colhead{f(H$\alpha$)} &
\colhead{EW(H$\alpha$)} &
 \colhead{Age (R08) \tablenotemark{a}} & 
 \colhead{Age\tablenotemark{b}} & 
\colhead{Mass} &
\colhead{ A$_V$} &
 \colhead{ R$_{\textnormal{H\sc{ii}}}$}\\
\colhead{} & 
\colhead{[erg s$^{-1}$ cm$^{-2}$]} & 
\colhead{\AA} & 
\colhead{Myr} & 
\colhead{Myr} & 
\colhead{$\msun$} &
\colhead{mag}&
\colhead{pc}}
\startdata
    SSC1& 5.14$\times10^{-14}$ (1\%)&3100.0(134.4)&$\leq 3.3$ &3.0 &  4.7$\times10^5$  & 0.73& 11.3(11.5)\tablenotemark{c}\\
    SSC2 & 3.34$\times10^{-14}$ (1\%)&2300.0(104.7)&$\leq 3.4$ &3.0 &  3.7$\times10^5$  & 0.65& 10.4(9.7)\tablenotemark{c}\\
    SSC3 & 5.77$\times10^{-15}$ (2.5\%)&787.0(53.4)&6.8(2.5)&7.0 &  7.1$\times10^5$  & 0.92&29.1(26.6)\tablenotemark{d}\\
    SSC4 & 2.42$\times10^{-15}$ (3.9\%)&176.0(10.7) &12.4(1.7)&  11.0  & 1.1$\times10^6$  & 0.20& 32.9(26.9)\tablenotemark{e}\\
    SSC5 & 2.29$\times10^{-15}$ (4.5\%)&84(4.4) & 15.1(2.3)&  13.0  & 2.9$\times10^6$  & 0.84&38.6(29.5)\tablenotemark{e}\\
    SSC6 & 6.34$\times10^{-16}$ (7.7\%)&187(22.5)&13.9(1.9) & 11.0  & 2.6$\times10^5$ &  1.08&20.6(21.49)\tablenotemark{e}\\
\enddata
\tablecomments{The first column shows the H$\alpha$ fluxes and the EW(H$\alpha$) determined in the present work. The errors are shown inside the brackets. The ages in the fourth and fifth column are obtained from the EW(H$\alpha$) by R08 and this work, respectively. The uncertainties associated to our age estimates are $\sim 1$ Myr. The following two columns show the values of the masses and visual extinctions produced by the model fit, at given age, of the full SEDs of each cluster. In the last column, we listed the predicted radii of the H{\sc ii}  around the clusters. See text for details.}
\tablenotetext{a}{Estimated ages by R08 comparing the observed EW(H$\alpha$) with SB99 predictions.}
\tablenotetext{b}{Ages estimated in this work using the re-estimated EW(H$\alpha$)  and Cloudy predictions for our models.}
\tablenotetext{c}{$n_H = 10^4$ cm$^{-3}$}
\tablenotetext{d}{$n_H = 10^3$ cm$^{-3}$}
\tablenotetext{e}{$n_H = 5\times10^2$ cm$^{-3}$}
\end{deluxetable*}

We also repeated the least-square fit done by R08, excluding the I band and IR data ({\it WFPC2}/F791W, and {\it NIC2} F160W, F187N, and F205W),  and using models with only stellar continua. In this case, age was treated as a free parameter. The fits with the smallest residuals are shown in Figure~\ref{spec} with red data points and red spectra. As already found by R08, the fit to the blue optical photometry using only stellar continua produced a quite evident displacement at wavelength longward of 8000 \AA. Nevertheless, when nebular continuum and line emission are included in the models, we obtained a good fit for all the clusters and in all the passbands. Inside the photometric errors, the best fitting model (black squares) is able to reproduce the SED shapes quite well. We noticed, however, a persisting small excess (between 0.2 and 0.5 mag above the models) at 8000 \AA \ (I band) which was observed in the bottom panel of Figure \ref{CCD} . We will discuss this feature in Section \ref{disc}.

We produced, finally, a third set of fits from UV to IR, using our models, where we let the age be a free parameter. The two youngest SSC1 and SSC2 had a best fitting age of 3 Myr in perfect agreement with the EW(H$\alpha$) determination. The SSC3, SSC4, SSC5, and SSC6 had best fitting ages lower than the values determined using EW(H$\alpha$): 6, 9, 9, and 7 Myr, respectively. However, investigating the 68.3 \% confidence levels, we constrained boundaries to the best fitting ages (Table  \ref{best_chi2}). We observe that inside the uncertainties produced by a free SED fit the solutions recovered using the observed EW(H$\alpha$) are in good agreement in all the cases except in SSC6 (marginally agreement). The estimated masses and extinction didn't change significantly (see Table \ref{best_chi2}). Therefore we do not discard any of the two results and consider the ages determined with the two methods as a valid range around the real age of the clusters.

\begin{deluxetable}{lccc}
\tablecolumns{4}
\tablewidth{0pt}
\tablecaption{Ages, masses, and extinctions of the clusters produced by the SED fit. In this case age is treated as a free parameter.\label{best_chi2}}
\tablehead{
\colhead{Ids} &
 \colhead{Age [Myr]} & 
\colhead{Mass [$10^5 \msun$]} &
\colhead{ A$_V$ [mag]}}
\startdata
    SSC1\tablenotemark{a}&3.0$^{+0.0}_{-2.0}$ &  4.7 & 0.73\\
    SSC2\tablenotemark{a} &3.0$^{+0.0}_{-2.0}$ &  3.7 & 0.65\\
    SSC3 &6.0$^{+1.0}_{-1.0}$ &  5.1  & 0.85\\
    SSC4 &  9.0$^{+2.0}_{-3.0}$  & 8.8  & 0.20\\
    SSC5 & 9.0$^{+6.0}_{-5.0}$  & 28.0  & 0.81\\ 
    SSC6 & 7.0$^{+2.0}_{-3.0}$  & 2.9 & 0.93\\
\enddata
\tablecomments{ }
\tablenotetext{a}{The best fitting ages, masses, and extinctions are the same as the one derived in the first set of fits for the youngest SSC1 and 2. See text for details.}
\end{deluxetable}

\subsection{Testing the validity of the models}
\label{tests}
Although our models can reproduce the shapes of the cluster SEDs, we need to test the consistency between the provided models and the physical conditions of the clusters.  The models include the full amount of emission flux coming from the photoionized gas. This assumption implies that all the ionized gas (e.g. the H{\sc ii} region) around the cluster is spatially contained in the used photometric aperture size. Using the nebular properties produced by the best fitting model, and cluster physical parameters published in previous works, we produced estimates  of the radii of the H{\sc ii} regions around the clusters, as expected from the model. The radii were computed using the formula:
\begin{equation}
R_{\textnormal{H{\sc ii}}} =  \left(\frac{3 Q_{Lyc}}{4\pi n_H^2\alpha_Bf}\right)^{1/3}
\end{equation}
under the assumption of a pure hydrogen cloud. 

\begin{figure}
\includegraphics[scale=0.48]{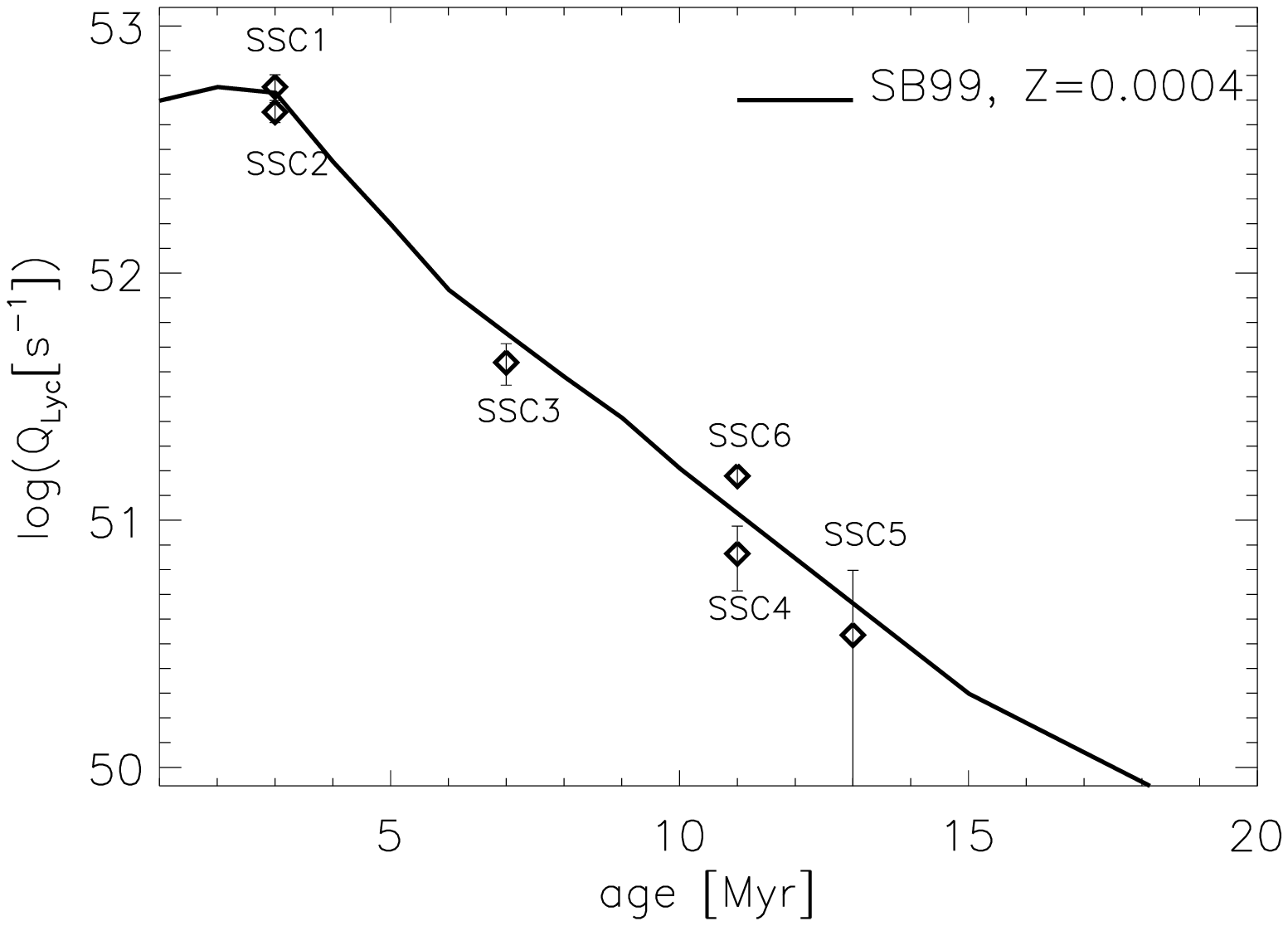}
\caption{Evolution in time of the $Q_{Lyc}$ produced by the Starburst99 model for a $10^6 \msun$ cluster. Overplotted the estimated $Q_{Lyc}$ of the six SSCs from the measured, extinction corrected f(H$\alpha$). The error bars show uncertainties of 25 \% on the estimations.}
\label{cont}
\end{figure}

The $Q_{Lyc}$ is the rate of ionizing Lyman continuum photons produced by  very young stars and changes as a function of cluster age. If the medium around the stars is clumpy or dusty, it is possible that a fraction of the ionizing photons do not ionized the gas. R08 discussed the possibility of a clumpy medium around the two youngest SSC1 and 2, and estimated a leakage of ionizing photons of $\sim 40$Ê\%. Our best fitting model had a covering factor $c=1$, implying that all the Lyman continuum photons, produced by the young massive stars, ionize the gas. To verify this result we compared the provided $Q_{Lyc}$ by the Starburst99 model with the values of $Q_{Lyc}$ derived from the measured H$\alpha$ fluxes (extinction corrected). The observed $Q_{Lyc}$ was estimated using the Relation (1) in R08, which produces lower limits to the real values. The f(H$\alpha$)  listed in our Table \ref{age-halpha}, were first corrected for the corresponding extinction values, and then converted into H$\beta$ luminosities. The resulting $Q_{Lyc}$ were scaled for the the mass of the model (10$^6 \msun$). We show the calculated values of the $Q_{Lyc}$ for the 6 SSCs in Figure \ref{cont}. For the two youngest clusters the estimated $Q_{Lyc}$ values are in agreement with the model predictions, and do not show any significant leakage of ionizing photons. We observe in the older clusters a possible escaping fraction of $20-30$ \%, consistent with the more evolved phases the clusters are experiencing. In general, we consider the assumption of $c=1$ a reasonable approximation. 

To estimate the radii of the H{\sc ii} regions of each cluster we used the $Q_{Lyc}$ predicted by the Starburst99 stellar continua, scaled for the mass of the cluster. In Table~\ref{age-halpha} we also show (into brackets) the values of the radii if the estimated $Q_{Lyc}$ from f(H$\alpha$) would be used instead.

Studies of massive star forming regions in our galaxy (\citealp{2004ASPC..322..329C}, \citealp{2002ApJ...564..827C}, among many others), have shown   quite complex environments with density gradient going from $n_H \approx 10^7$ cm$^{-3}$ in pre-stellar cores to diffuse surrounding regions with $n_H \approx 10^2$ cm$^{-3}$, and ultra-compact  H{\sc ii} regions ($n_H \approx 10^4$ cm$^{-3}$), all coexisting in the same system.  \citet{2006ApJ...638..176T} presented observational evidence of very compact H{\sc ii} regions in the position of the two young SSC1 and SSC2. On the other hand, using two-dimensional spectroscopy, \citet{2006A&A...459...71I} found values of $n_H$ across the SSCs  of a few hundreds and EW(H$\alpha$) below 1500 \AA \ even in the brightest H$\alpha$ regions of the galaxy (SSC1+2). We noticed, however, that the spectra are produced in regions approximate 10 times bigger (0.52"$\times$0.52") than the ones used here (radius of $\sim$0.15"), causing a smoothing of the values over low density and diffuse regions around the clusters. Recently using continuum radio data, \citet{2009AJ....137.3788J} estimated for SSC1 and 2 a $n_H\sim 4-7 \times 10^3$ cm$^{-3}$ in agreement with the value of $n_H=10^4$ cm$^{-3}$ we derived from the fitting model. 

In order to estimate the radii of the H{\sc ii} regions for the two youngest clusters we used $n_H=10^4$ cm$^{-3}$. In the previous section, the model with lower $n_H$ produced also a good fit to the data. By keeping all the other parameters fixed, we checked whether a lower value of $n_H=100$ cm$^{-3}$ produced very different flux values in the optical I band and in the NIR wavebands for ages between 6 and 15 Myr. We found that the difference between the two models was less then 1\%. Finally, we decided to use $n_H=10^3$ cm$^{-3}$ for SSC3, which has an intermediate age, and a density of  few hundreds, for the older three SSCs. 

The filling factor, f, was fixed to 0.01. We set the recombination coefficient, $\alpha_B$, for Case B recombination \citep{2006agna.book.....O}, to 1.43$\times 10^{-13}$ at an electron temperature of 20000 K  (determined by \citet{1997ApJ...476..698I} from optical emission lines of the galaxy).  

The values we estimated are listed in Table~\ref{age-halpha}. We derived quite small sizes for the H{\sc ii} regions of SSC1 and 2, due to the high value of $n_H$ of the photoionized cloud surrounding the recently born clusters. The sizes increase for lower values of $n_H$. This behavior is expected since the ionizing photons travel farther in less dense environments. Nevertheless we found consistency between the assumptions made in the models and the aperture used by R08 to make the photometric measurement.

\section{Discussion}
\label{disc}
\subsection{The origin of the IR excess in the SSCs}

R08 presented evidence of a flux excess in the red optical and NIR wavebands. The excess at $\sim$ 0.8 $\mu$m was attributed to the ERE phenomenon. The flux excess at wavebands between 1.6 and 2.1 $\mu$m was considered to have two different origins. For the youngest clusters SSC1 and SSC2,  hot dust ($\sim800$ K) emission was advocated to explain the rise of the IR continuum. For the remaining SSC3, 4, 5 and 6, they showed that the infrared colors  were consistent with red super giants (RSGs) which mainly contribute to the NIR light at ages above 7 Myr. 

We propose a new interpretation of the flux excess in the SBS 0335-052E SSCs. We have shown that the inclusion of the emission from photoionized gas in the models can well reproduced the colors (Figure~\ref{CCD}) and the SED shape (Figure~\ref{spec}) of all the SSCs from UV to NIR. The ages found using the new estimated EW(H$\alpha$) are very similar to the one previously determined from R08 and confirm the general picture of the star formation propagating from North to South (\citealp{2006ApJ...638..176T}; R08). Moreover previous spectro-photometric studies of the galaxy from optical to infrared confirmed the presence of dominant nebular contribution (\citealp{2000A&A...363..493V}, \citealp{2001A&A...378L..45I}). \citet{2000A&A...363..493V} estimated that more than 45 \% of flux in the NIR spectra of the SSC1+2 and SSC4+5 regions are due to nebular continuum and emission. They also found from H$\beta$/Br$\gamma$ ratio a visual extinction of $A_V \approx 0.73$, which is in good agreement with the values we found from the SED fits. Finally, they interpreted the red $H-K$ color observed globally in the galaxy as due to a hot dust component ($\sim$670 K). However due to the very bright Br$\gamma$ emission and low extinction observed across the two knots (SSC1+2 and SSC4+5), they concluded that even if there may be a large quantity of dust in the galaxy, it is not located inside the SSCs. Due to the very young ages of SSC1 and SSC2, the absence of any treatment of the nebular emission, and the very high temperature of the dust advocated to explain the red excess in R08, make this explanation difficult to reconcile with the previously found observational properties. We showed that the integrated fluxes of these two clusters can easily be fitted from UV to NIR with self-consistent models including both stellar and nebular components. 

For the remaining clusters R08 showed that the colors are compatible with the presence of RSGs, which are important in the NIR range at cluster ages around $7-10$ Myr. 

However, \citet{2006A&A...454..119P}  found for SSC3 evidence of Wolf-Rayet stars (WRs) in the optical spectrum. The presence of WRs is a very transient phenomenon and Starburst99 models predict that they last around 1 Myr at a cluster ages of around 6 Myr. This age is in agreement with our estimate for SSC3. It is likely that SSC3 is still too young to show contribution due to RSGs. 

In the NIR spectrum of the knots SSC4+5, \citet{2000A&A...363..493V} found no evidence of CO absorption typical of RSGs and important if the nebular emission does not contribute any longer \citep{2003ApJ...596..240M}. We have to mention that the regions where SSC4 and SSC5 are located are very complex due to the proximity of very young star forming sources. It is possible that the spectra presented by \citet{2000A&A...363..493V} are indeed dominated by these very young star forming regions. 

 However we detect H$\alpha$ emission at the location of SSC3, SSC4, SSC5, and SSC6, which reveal that the emission from the photoionized gas is not gone yet. We have showed in Section \ref{sec:mod} that the fraction of nebular emission contributing in the NIR filters (bottom panel of Figure \ref{model2}) is still considerable and is strictly connected with the metallicity of the gas. The same models but at higher metallicity (i.e. solar metallicity) show that the nebular phase is over at around 6 Myr. Since  SBS 0335-052E is among the known lowest metallicity nearby galaxies, we expect to see such prolonged contribution.

Both R08 and T09 analysed the Pa$\alpha$ NICMOS data available for SBS 0335-052E, reaching discordant results regarding the Pa$\alpha$ emission in the SSC3, SSC4, SSC5, and SSC6. T09 reported a none detection of Pa$\alpha$ emission in these clusters, showing a dearth of ionised gas. On the other hand, the values reported by  R08 (listed in their Table 3) are close to the ones predicted by the H$\alpha$/Pa$\alpha$ ratio considering our derived extinction.

\subsection{What is the cause of the I band excess?}

In the previous section, we noticed that the six clusters have the observed flux in the I band (F791W) that sit above the best model predictions, with a discrepancy of $0.2-0.5$ mag. It is likely that the flux excess in this filter is due to another mechanism than photoionized gas. The presence of a strong UV radiation field fuelled by the very young massive stars and large quantities of dust distributed in the galaxy and close to the SSCs can originate the ERE phenomenon as already proposed by R08. 

An other possible explanation for this small systematic offset could be related to the low resolution power of the F791W data. The coarse pixel scale of the {\it WFPC2}/WF3 chip ($0.1"$) used to capture the F791W image and the small aperture radius used (0.15") means that the aperture correction is significantly more uncertain than for the remainder of the UV/optical data which was obtained with ACS.  A significant error in the F791W aperture correction would  lead to the SSCs lying systematically above or below the best fit models. The accuracy of the encircled energy distribution determination for a  0.15" radius aperture is no better than 0.07 magnitudes \citep{1995PASP..107..156H}. While this is smaller than the observed excess ($\sim 0.3$mag) it cannot be excluded that an uncertainty in the aperture correction contributes to the apparent offsets. The clusters are poorly sampled in the image and this could produce higher uncertainties associated with the derived aperture correction.

\section{Conclusions}
\label{conc}

We tested our models which include stellar continua and a self-consistent treatment of the photoionized gas to verify the impact of the nebular component on the total flux of the young star clusters. Using new measurements of the EW($\alpha$) we determined the ages of the 6 bright SSCs in SBS 0335-052E. The ages were used to produced new fits to the whole SED of the clusters using our custom-designed evolutionary models. With the contribution to the integrated fluxes from the photoionized gas, we can fairly reproduce the SED shape in all the six clusters. This new proposed explanation is also supported by previous numerous analysis of the starburst regions in the galaxy. 

We verified whether the assumptions made by our models are in agreement with the observed physical properties of the clusters and in general, of the starburst environment. We also tested the condition that the whole ionized region contributing to the total integrated fluxes is contained in the aperture size used to do photometry. We found that the size of the emitting H{\sc ii} regions around the clusters are reasonably smaller than the photometric radii.

We finally noticed a persistent flux excess at 0.8 $\mu$m, impossible to reconcile with our spectral evolutionary models. We consider, in agreement with R08, the possibility that dust photoluminescence, typically observed in H{\sc ii} regions around massive star forming complexes, can contribute to the integrated flux transmitted by the {\it WFPC2}/F791W filter. 

 While nebular emission seems to be the origin of the red excess in the two dwarf galaxies SBS 0335-052E (this work) and NGC 4449 \citep{R2009,2008NGC4449R}, this is not the case for the red excess observed in the young star clusters of Haro 11 \citep{Adamo10}, which remains unresolved. The access to X-shooter data for the SBS 0335-052E SSC1+2 and few of the clusters with strong red excess in Haro 11, would represent a fundamental source of information to disentangle between the different causes of the red excess, and quantify the real amount of flux produced by nebular continuum and emission.

\acknowledgments

We thank Amy Reines for valuable discussions and suggestions made on this work. Nate Bastian and Brent Groves are gratefully acknowledged for the useful comments made on this manuscript. A.A., G.\"O., and E.Z. acknowledge support from the Swedish Research council. E.Z. acknowledges research grants from the Swedish Royal Academy of Sciences. G.\"O. is a  Royal Swedish Academy of Sciences research fellow, supported from a grant from the Knut and Alice Wallenberg foundation. G.\"O. and E.Z. also acknowledge the Swedish National Space Board. M.H. acknowledges the support of the Swiss National Science Foundation. This research has made use of the NASA/IPAC Extragalactic Database (NED) which is operated by the Jet Propulsion Laboratory, California Institute of Technology, under contract with the National Aeronautics and Space Administration.





\clearpage

\end{document}